\documentclass[11pt]{article}
\usepackage[margin=2.5cm]{geometry}
\usepackage{color}
\usepackage{xcolor}
\usepackage{graphicx}
\usepackage{amsmath}
\usepackage{amsfonts}
\usepackage{soul}
\usepackage[normalem]{ulem}
\usepackage{lipsum}
\usepackage{footnote}
\usepackage{bibentry}
\usepackage{float}
\usepackage{authblk}
\usepackage{booktabs}
\usepackage[hidelinks]{hyperref}
\usepackage{silence}
\usepackage{cite}

\begin{document}

\title{Non-destructive sub-surface inspection of multi-layer wind turbine blade coatings by mid-infrared Optical Coherence Tomography}

\author[1]{Lapre C.}
\author[1,2]{Petersen C. R}
\author[3]{Nielsen P.}
\author[3]{Wulf T.}
\author[4]{Bech J. I.}
\author[4]{Fæster S.}
\author[1,2,5]{Bang O.}
\author[1,2]{Israelsen N. M}
\affil[1]{DTU Electro, Technical University of Denmark, 2800 Kgs. Lyngby, Denmark}
\affil[2]{NORBLIS ApS, Virumgade 35 D, 2830 Virum, Denmark}
\affil[3]{Siemens Gamesa Renewable Energy A/S, Assensvej 11, 9220 Aalborg, Denmark}
\affil[4]{Department of Wind and Energy Systems, Technical University of Denmark, Roskilde, Denmark}
\affil[5]{NKT Photonics, Blokken 84, Birkerod 3460, Denmark}
\date{}

\maketitle

\begin{abstract}
Non-destructive inspection (NDI) is useful in the industrial sector to ensure that manufacturing follows defined specifications, reducing the quantity of waste and thereby the cost of production. In response to climate change and the need to reduce carbon footprints, there is a growing demand for NDI solutions that are primarily non-contact, as well as non-destructive, fast, and easily integrated. Optical Coherence Tomography (OCT), a well-known diagnostic technique in medical and biological research, is increasingly being used for industrial NDI. In the mid-infrared (MIR) wavelength range, OCT can be used to characterise parts and defects not possible by other industry-ready scanners, and enables better penetration than conventional near-infrared OCT. 

In this article, we demonstrate NDI of wind turbine blade (WTB) coatings using an MIR OCT scanner employing light around 4~$\mu$m from a supercontinuum source. We inspected the top two layers of the coating (topcoat and primer) in two different samples. The first is to determine the maximum penetration depth, and the second one is to emulate defect identification. The results of our study confirm that MIR OCT scanners are highly suitable for coating inspection and quality control in the production of WTBs, with performance parameters not achievable by other technologies.
\end{abstract}

 \section{Introduction}

In recent years, the development of renewable energy sources has greatly intensified, in part due to the ongoing transition away from fossil fuels. For countries like Denmark, where coastal winds are powerful and constant, developing new offshore wind turbines is essential. However, the typical lifespan of wind turbine technology is limited to around 20 years, or up to 25 years in optimal conditions, which is mainly due to erosion on the blades \cite{Lichtenegger2020,Beauson2022}. This research is part of a European project aimed at developing new offshore WTBs with extended lifetimes and reduced environmental impact during manufacturing \cite{TURBO_project}. \newline

One of the critical challenges in manufacturing WTBs is achieving durable and high-quality coatings. These coatings are vital for protecting the blade's internal structure and mitigating delamination effects \cite{Slot2015,Mishnaevsky2021,Faester2021,Mendonca2023}. Currently, the most common method used for testing the coating quality in production is a so-called ``dolly test'', a destructive method where the coating layer is removed in a region of interest and then characterised. This approach is inherently destructive, limited in scope, and cannot provide a comprehensive evaluation of the entire blade.\newline

In industrial contexts, Non-Destructive Inspection (NDI) techniques play a crucial role in ensuring that materials, samples, or parts follow manufacturing specifications. Vibration analysis or acoustic emission techniques allow manufacturers to detect early damage, or crack propagation; such as Laser Doppler vibrometry (LDV) which can detect cracks as small as around $25$~mm, as demonstrated in recent studies \cite{Yang2018,Asokkumar2021,Zabihi2024}. Thermography is easy to use, achieves an average penetration depth of $3$~cm, and works well with composite materials like GFRP \cite{Meola2010,Liu2017,Bernegger2020} but the user has to take care about unwanted thermal damage, and it only provide a surface projection. For deeper inspections, radiographic methods (e.g., X-rays or $\gamma$-rays), and microwave (super high frequency or terahertz frequency) inspection are effective methods for detecting internal faults in blades with the advantage of requiring no contact for inspection. However, for large-scale inspections, such as an entire WTB, the transverse or lateral resolution is often too low and impracticable for coating characterisation, and the radiation emitted by radiographic methods is dangerous \cite{Mishnaevsky2021,Faester2021,Li2016,Taraghi2019,ShinYee2024}. Table~\ref{Table_1} presents a summary of the NDI technologies available for WTB inspection in the production \cite{GarciaMarquez2020,Aminzadeh2023}. 
\newline

\begin{table}[H]
\resizebox{\textwidth}{!}{%
\begin{tabular}{|l|l|l|l|l|l|}
\hline
\textbf{NDI method}& \begin{tabular}[c]{@{}l@{}} \textbf{Surface }\\ \textbf{inspection}\end{tabular} & \begin{tabular}[c]{@{}l@{}}\textbf{Volume}\\ \textbf{inspection}\end{tabular} & \begin{tabular}[c]{@{}l@{}}\textbf{Penetration} \\ \textbf{depth}\end{tabular}& \textbf{Defect depth resolution}& \begin{tabular}[c]{@{}l@{}}\textbf{Image} \\ \textbf{output}\end{tabular}\\ \hline
Thermography& Yes&  Volume projection & $\sim$30 mm   &   few millimetres          & \begin{tabular}[c]{@{}l@{}}2D : top view\\ projection\end{tabular} \\ \hline
LDV& Yes & Yes& few centimetres &  $\sim$ 25 - 30 mm & \begin{tabular}[c]{@{}l@{}}2D : top view\\ projection\end{tabular} \\ \hline
Radiographic CT* & Yes  & Yes & \begin{tabular}[c]{@{}l@{}}few micrometres \\ to metres\end{tabular}&  $\sim$ 0.1 - 20 mm & 3D \\ \hline
\begin{tabular}[c]{@{}l@{}}Ultrasonic and\\ vibration\end{tabular}& Yes & Yes & few milimetres &  $\sim$ 0.1 - 30 mm   & 3D \\ \hline
\begin{tabular}[c]{@{}l@{}}Microwave \\ (super high frequency \\ and terahertz frequency) \end{tabular}& Yes & Yes & $\sim$1-10 mm &
$\sim$ 90 - 150 mm & \begin{tabular}[c]{@{}l@{}}2D : top view\\ projection\end{tabular} \\ \hline
NIR OCT ($\sim$1.3~$\mu$m)  & Yes & Yes & 
10 - 100 $\mu$m: & $\sim$ 1 - 15 $\mu$m  & 3D\\ \hline
MIR OCT ($\sim$4~$\mu$m) \textdagger & Yes   & Yes  & $\sim$ 400 $\mu$m:&  $\sim$ 7 - 30 $\mu$m & 3D \\ \hline
\end{tabular}}
\caption{Non-exhaustive summary of NDI technologies for characterisation of multilayer coating and or GFRP (cf. Fig.~\ref{Figure_1}).\\
*  X-rays or $\gamma$-rays, not feasible on large sample. \\
\textdagger~Our MIR OCT system}
\label{Table_1}
\end{table}

Another imaging technique well-known in the biomedical field for several decades, but less common in the industrial sector, is OCT \cite{Fujimoto2000,Drexler2001}, based on combined laser scanning and sub-surface light echoes deduced from interference signals. This technology has the advantage of being fast and suitable for characterising organic materials, and achieving ultra-high depth resolution around $1$-$50~\mu$m and a penetration depth around $1$-$2$~mm, depending on the wavelength of the light that is being used. Mid-infrared (MIR) OCT scanning, i.e. above $2$-$3~\mu$m, has already demonstrated its capability for NDI of marine coatings \cite{Petersen2021}, WTB coatings \cite{Petersen2023,Lapre2024_2}, paper quality inspection \cite{Hansen2022}, credit card inspection \cite{Israelsen2019}, and ceramic inspection \cite{Zorin2022,Lapre2024}. The main disadvantage of this technology is its limited penetration depth, but for the top surface of blade coatings, including the topcoat and primer, a penetration depth of several hundred micrometres is sufficient. This article demonstrates non-contact, non-destructive sub-surface inspection of defects in WTB coatings, up to a depth of $\sim360~\mu$m, of the combined topcoat and primer, using a MIR OCT scanner employing a supercontinuum laser source around $4~\mu$m centre wavelength.

\section{Materials and methods}

\subsection{The blade samples} \label{the_blade_sample}

\begin{figure}[H]
\centering\includegraphics[width=15cm]{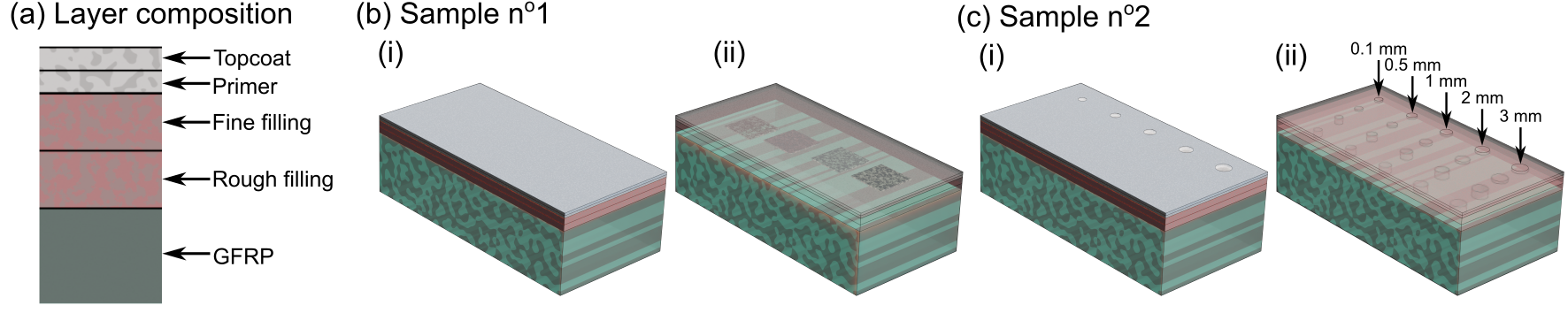}
\caption{(a) Schematic of the layer composition of the samples visualised in (b,c). (b) Sample with aluminium foil patches placed at the interfaces between layers. (c) Sample with punch defects in different layers. For (b-c), (i) and (ii) show opaque and transparent 3D illustrations of the samples.} 
\label{Figure_1}
\end{figure}

Two sets of samples were tested, denoted $\mathrm{n^{o}1}$ and $\mathrm{n^{o}2}$. Figure~\ref{Figure_1}(a) shows a schematic of the layer composition above the GFRP, and Fig.~\ref{Figure_1}(b-c) presents 3D illustrations of both samples. During the process of fabrication, the layers are subsequently applied and dried in a controlled environment. Each layer is sanded before the application of the next one. The materials used for each layer are as follows:

\begin{enumerate}
    \item Topcoat CWind UHS Topcoat + Relest Hardener PUR 307 (CLEAR) + Relest Thinner PUR 307.
    \item Primer : Carboline Windmastic$^{\mbox{\scriptsize{\textregistered}}}$400 FC Primer (PART A and B) + Carboline Thinner no.25.
    \item  Fine filling (Awl 8020) : Awlfaire Pumpable Base + Awlfaire Pumpable Converter.
    \item Rough filling (Awl 8200) : Awlfaire LW White Base + Awlfaire LW Fast Converter.
\end{enumerate}

Sample $\mathrm{n^{o}1}$, presented in Fig.~\ref{Figure_1}(b), was used to benchmark the penetration depth of the MIR OCT scanner and the refractive index of the layers. Different aluminium foils were inserted in all the layer interfaces as seen in the transparent 3D view in Fig.~\ref{Figure_1}(b-ii). 

Sample $\mathrm{n^{o}2}$, presented in Fig.~\ref{Figure_1}(c), was made to mimic the defect that can appear inside the coating during the manufacturing process, in our case, it was a hole defect. Fig.~\ref{Figure_1}(c-i) and (c-ii) show an opaque and a transparent 3D visualisation of this sample. For each layer, 5 holes with different diameters were made by punching the paint in staggered rows with metallic needles with different diameters: $0.1$~mm, $0.5$~mm, $1$~mm, $2$~mm, and $3$~mm.  \newline

The layer thicknesses of both samples are summarised in Table~\ref{Table_2_3}. The wet film thicknesses were provided by Siemens Gamesa, while the dry film thicknesses were calculated from the datasheet of the paint, the MIR OCT measurements, and the X-ray CT measurements. In a previous study \cite{Petersen2021}, it was demonstrated that MIR OCT has the capacity to measure the thickness of wet paint film and follow the shrinking of the paint during curing; this corroborate the difference of thickness measured from the OCT and X-ray CT.

\begin{table}[H]
\resizebox{\textwidth}{!}{
\begin{tabular}{l|c|c|c|c|}
\cline{2-5}
& \multicolumn{4}{|c|}{Layers of sample $\mathrm{n^{o}1}$}\\
\cline{2-5}
& Topcoat [$\mu$m] & Primer [$\mu$m] & Fine filling [$\mu$m] & Rough filling [$\mu$m]\\ 
\hline
\multicolumn{1}{|l|}{Wet film}& 210  & 290 & 500 & 1050 \\
\hline
\multicolumn{1}{|l|}{Dry film calculated with datasheet} & $\sim 149$  & $\sim 255$ & 500 & 1050 \\
\hline
\multicolumn{1}{|l|}{Dry film measured with OCT above foil
}  & $\sim 144\pm 3.3$  & $\sim 219\pm 2.6$ & unknown & unknown \\
\hline
\cline{2-5}
\cline{2-5}
& \multicolumn{4}{|c|}{Layers of sample $\mathrm{n^{o}2}$}\\
\cline{2-5}
& Topcoat [$\mu$m] & Primer [$\mu$m] & Fine filling [$\mu$m] & Rough filling [$\mu$m]\\ 
\hline
\multicolumn{1}{|l|}{Wet film}& 170  & 190 & 500 & 2000 \\
\hline
\multicolumn{1}{|l|}{Dry film calculated with datasheet} & $\sim 120$  & $\sim 166$ & 500 & 2000 \\
\hline
\multicolumn{1}{|l|}{Dry film measured with OCT} & $\sim 120 \pm 3.3$  & $\sim 123 \pm 2.6$  & unknown & unknown \\
\hline
\multicolumn{1}{|l|}{Dry film measured with X-ray}& \multicolumn{2}{c|}{ $\sim 240\pm7.5$}  & $\sim 338\pm7.5$ & $\sim 1539\pm7.5$\\
\hline
\end{tabular}}
\caption{ Wet and dry film layers thicknesses in micrometres for the ``foil sample'' (sample $\mathrm{n^{o}1}$), and for the ``defect sample'' (sample $\mathrm{n^{o}2}$).}
\label{Table_2_3}
\end{table}

\subsection{The OCT scanner} \label{mir_oct_section}

\begin{figure}[H]
\centering\includegraphics[width=15cm]{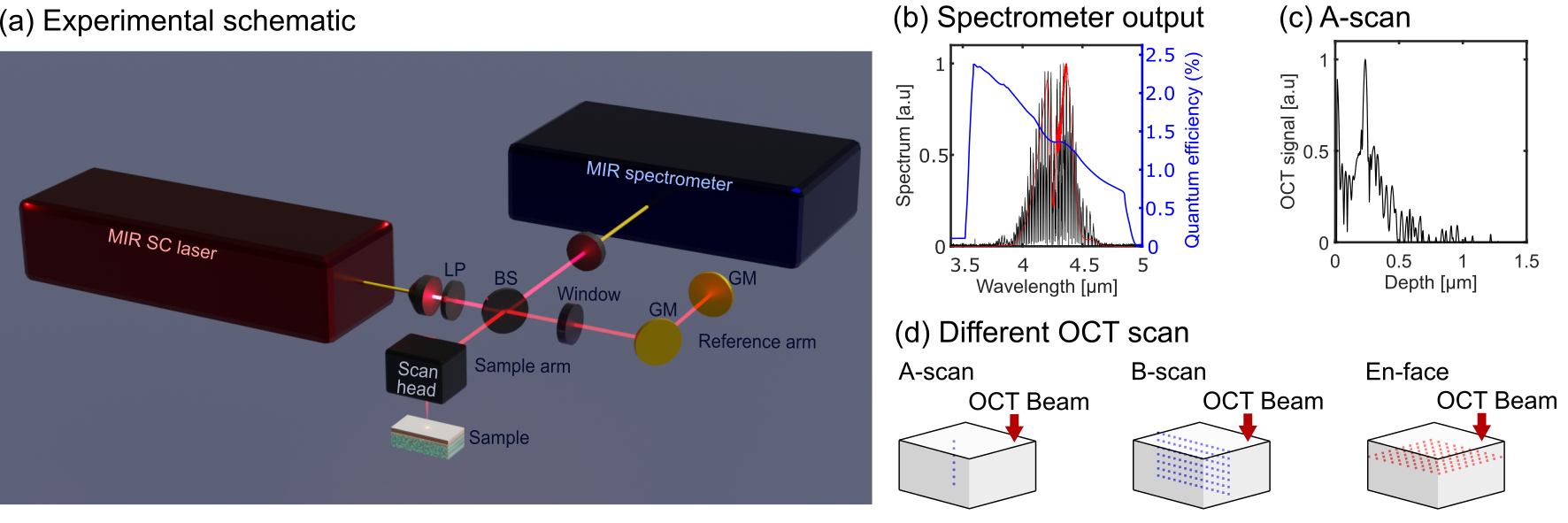}
\caption{(a) 3D representation of the MIR OCT system,  SC: supercontinuum, LP: 3.2~$\mu$m long pass filter, BS: beam splitter, GM: gold mirror, Window: dispersion compensation. (b) Reference spectrum in red, measured interference spectrum in black, upconversion efficiency in blue, and (c) the corresponding calculated A-scan. (d) Schematic of the different OCT scans used in this article. } 
\label{Figure_2}
\end{figure}

The MIR OCT scanner is presented in Fig.~\ref{Figure_2}(a). The light source is an in-house fabricated MIR supercontinuum (SC) source covering a spectrum from 1 to 4.6~$\mu$m \cite{woyessa2021power} coupled to a Michelson interferometer. Due to the range of the spectrometer and to minimise the amount of unnecessary power, there is a long-pass filter at the input of the interferometer, which removes light at wavelengths shorter than 3.2~$\mu$m (cf. Fig.~\ref{Figure_2}). At the output of one arm of the interferometer, the scanning is achieved using a two-axis silver-coated mirror galvanometric scanner coupled to an achromatic germanium lens with a focal length of 30~mm. The average power on the sample arm was about 20~mW.\newline 

The lens inside the scan head introduces chromatic dispersion, which occurs due to a variation in the speed of light depending on the wavelength as it travels through a material. In response, a germanium window was inserted in the reference arm to compensate the dispersion, the residual dispersion mismatch was compensated numerically \cite{Wojtkowski2004}. The system achieved an axial (depth) resolution of  $\sim$ 9.74~$\mu$m, $\sim$ 6.54~$\mu$m in topcoat, and $\sim$ 5.29~$\mu$m in primer, and a transverse spatial resolution of $\sim$ 22~$\mu$m, determined by using a standard USAF-1951 target. The maximum sensitivity was 60 dB and the 6 dB sensitivity roll-off depth was 1.9 mm in air for a 0.3~ms A-scan integration time. For more information about the MIR OCT system, see \cite{Israelsen2019,Israelsen2021}. To compare with more commonly available OCT systems, the measurement was also made with a near-infrared (NIR) OCT system with a centre wavelength around $1.3~\mu$m (cf. Fig.~\ref{Figure_3}(c-d)), which is described in \cite{Israelsen2017,Israelsen2018}.

\subsection{Scanning parameters}\label{scanning_parameters}

The scanning parameters for the characterisation of the sample $\mathrm{n^{o}1}$ were the following, 500 cross-sectional images (B-scans) composed of 500 depth scans (A-scans). For sample $\mathrm{n^{o}2}$, 1000 B-scans composed of 1000 A-scans were used. Both sets of scan parameters covered an area of $\sim$5.73$\times$5.73 mm$^2$, with each A-scan and B-scan being separated by $\sim$11.49~$\mu$m for the sample $\mathrm{n^{o}1}$, and by $\sim$5.73~$\mu$m for the sample $\mathrm{n^{o}2}$. Each A-scan took 2500~$\mu$s to be acquired. As the lens adds curvature to the cross-sectional image, a flattening post-processing procedure was applied for correcting this effect. The different types of OCT scans presented in this article are illustrated in Fig.~\ref{Figure_2}(d). 

\subsection{Thickness estimation}

A beam propagating inside a turbid medium is attenuated as a result of scattering and absorption. In the case of light propagating in a homogeneous medium, the irradiance $L(z)$ $[\mathrm{W~cm^{-2}}]$ of the beam follows Beer-Lambert's law: 

\begin{equation}
    L(z) = L_0 e^{-\mu z}
    \label{BL_law}
\end{equation}
with $L_0$ the incidence irradiance and $\mu$ the attenuation coefficient \cite{Vermeer2013,Li2020}. Extraction of the attenuation coefficient from OCT measurements is done by fitting the exponential function from equation~\ref{BL_law} to the OCT signal as a function of depth, the A-scan. As OCT images are most commonly depicted on a logarithmic signal form, an exponential decay appears as a linear decay in the OCT A-scans. The presence of noise and speckles inside our data requires averaging over more than a dozen A-scans to enhance the precision in determining the different linear signal decay rates in the different layers \cite{Bashkansky2000,Desjardins2007}.\newline

We developed a primitive automatic interface calculation program based on fitting the attenuation coefficient of the different coating layers. To develop this algorithm, we first calculated the \textit{objective fits} for the decay of the different layers outside any intentional defect region. As the coating layers are not totally homogenous and to reduce noise, speckle, and potential non-intentional defect fluctuation, we calculated a smooth B-scan from an the average of 100 neighbouring B-scans. Then, we calculated an averaged A-scan profile on the entire size of the smooth B-scan, and the decay rate \textit{objective fits} were determined. \newline

After the calculation of this \textit{objective fit}, from an area (hopefully) without defects, we can then run the automatic interface recognition program in the interested area. It is possible to run the program with no averaging B-scan, but to enhance the automatic recognition of each interface, we averaged 10 neighbouring B-scans around the interested B-scan and then applied the algorithm on it. The algorithm follows the protocol outlined below:

\begin{enumerate}
\item Calculate an average A-scan with X neighbouring A-scan.
\item Identify the surface position inside the average A-scan.
\item Define an initial range of interest from the surface position to the end of the A-scan.
\item Calculate the linear fit of the A-scan in this range.
\item If the fit doesn't correspond to the \textit{objective fit} value.
    \begin{enumerate}
        \item Reduce the range of interest.
        \item Calculate the linear fit.
        \item Compare with the objective value. 
        \item If the fit doesn't correspond to the \textit{objective fit} value, then repeat \textit{5} until it does or the range becomes zero.
    \end{enumerate}
\item Proceed with \textit{3}, \textit{4}, and \textit{5} for the primer layer, and the fine filling.
\item The noise fit is the same as the one calculated for the objective.
\item In order to detect the interface, the crossing point between the topcoat and the primer fit, and primer and noise fit is calculated. 
\item Repeat steps from \textit{1} to \textit{8} for the rest of the average A-scan.
 \end{enumerate}

At the end, the program provides a B-scan with different plot lines, which shows the estimation of the surface, and each layer interface.\newline

\section{Results}\label{section_results}
Evaluation of the capability of MIR OCT for NDI of WTBs follows a number of steps. First, a comparative NIR OCT and MIR OCT scanner benchmarking of the penetration depth is presented. Secondly, imitated WTB defects are evaluated for MIR OCT against X-ray CT images. In continuation, coating layer delineation is presented, and afterwards an example of sub-surface additional defect categorisation is finally given.

\subsection{Benchmarking against NIR OCT} 

\begin{figure}[H]
\centering\includegraphics[width=15cm]{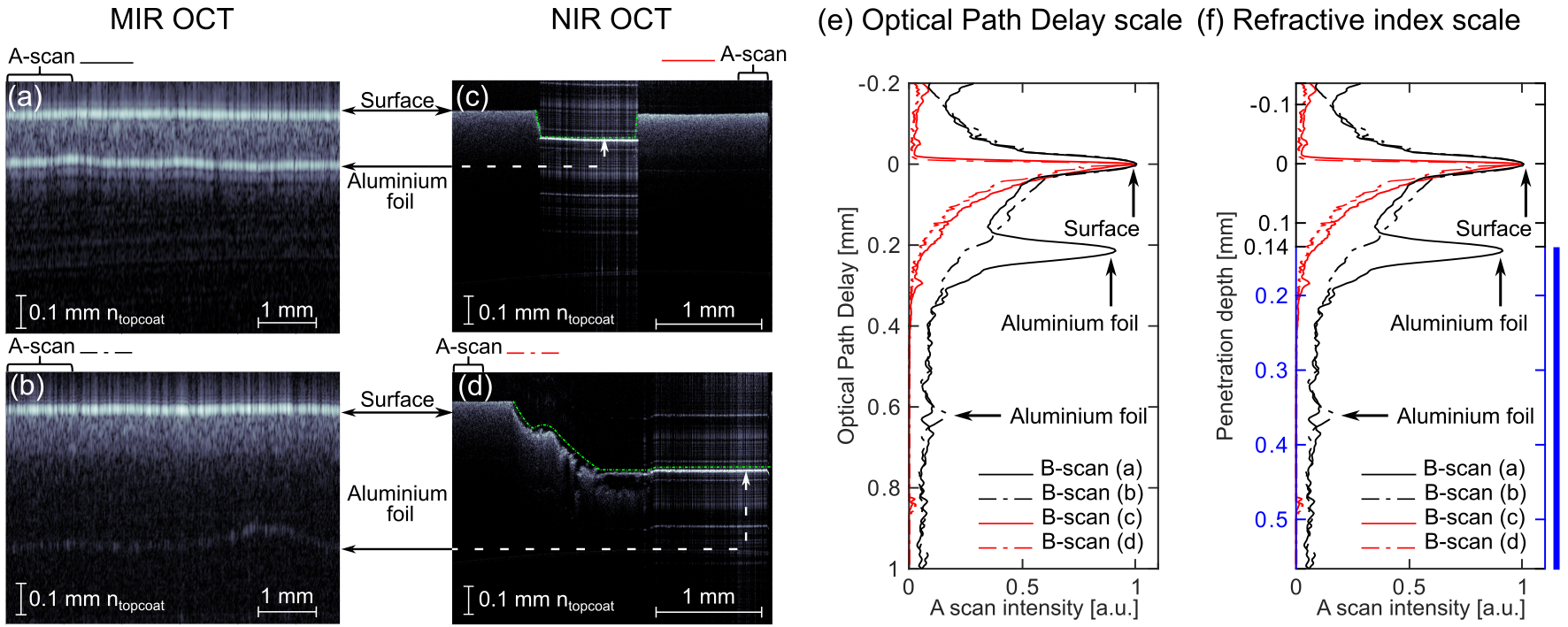}
\caption{(a-b) Comparison of penetration depth between B-scans in MIR OCT ($4~\mu$m centre wavelength) and (c-d) in NIR OCT ($1.3~\mu$m centre wavelength). The dashed green line in (c-d) shows the part where the coating was intentionally removed for the NIR OCT measurement to find the presence of aluminium foil. For each B-scan, the physical distance in depth was calculated with $n_{\mathrm{topcoat}} \simeq 1.49$. \\ (e-f) Comparison of the A-scan average of each B-scan is depicted in (a-d). (e) Present the results with the Optical Path Delay ($OPD$) scale and (f) scale with $n_{\mathrm{topcoat}} \simeq 1.49$ (from 0 to 0.14~$\mu$m ) and then scale with $n_{\mathrm{primer}} \simeq 1.84$ (blue scale from 0.14~$\mu$m).}
\label{Figure_3}
\end{figure}

Sample $\mathrm{n^{o}1}$,  containing interface aluminium foil markers, was used to assess the penetration depth of MIR OCT against conventional NIR OCT. The MIR OCT scanner could penetrate both the topcoat and the primer, detecting a clear reflection from the surface and the top two aluminium foils between the topcoat and primer and between the primer and the fine filling. This is observed in Fig.~\ref{Figure_3}(a,b) where respective layer interface foil signals are mapped out. These signals present a bright depth boundary resembling the signal of a mirror below the bright surface signal. The foil between fine filling and rough filling was not detected with the MIR OCT system, presenting a penetration depth limit in terms of layer interfaces. For NIR OCT scan tests, penetration performance was expected to be inferior due to the significant scattering. For this reason, in a small area, the coating was entirely removed until the foils in order to locate them in the B-scans. The part where the coating above aluminium foil was removed is highlighted by a dashed green line in Fig.~\ref{Figure_3}(c,d).\newline

As $n =  OPD/L$, with $OPD$ the Optical Path Delay, $L$ the physical distance, and $n$ the refractive index of the media, it is possible to calculate the refractive index of the topcoat and the primer. NIR OCT B-scan and the MIR OCT B-scan were use to calculated both refractive index. The physical distance of topcoat and primer were measured from the NIR OCT B-scan, from the surface of the sample until the surface of the aluminium foil in the area were the coating was removed (cf. green dashed line in Fig.~\ref{Figure_3}(c,d)), $L_{\mathrm{topcoat}} = 144~\mu$m and $L_{\mathrm{topcoat+primer}} = 363~\mu$m. Then the $OPD$ were extracted from the  MIR OCT B-scans (Fig.~\ref{Figure_3}(a,b)), $OPD_{\mathrm{topcoat}} = 215~\mu$m, and  $OPD_{\mathrm{topcoat+primer}}= 618~\mu$m.

The refractive index ot topcoat and primer are :
\begin{equation}
    n_{\mathrm{topcoat}} = \dfrac{OPD_{\mathrm{topcoat}}}{L_{\mathrm{topcoat}}} \simeq 1.49
\end{equation}

\begin{equation}
    n_{\mathrm{primer}} = \dfrac{(OPD_{\mathrm{topcoat+primer}}-OPD_{\mathrm{topcoat}})}{(L_{\mathrm{topcoat+primer}}-L_{\mathrm{topcoat}})} \simeq 1.84
\end{equation}

Figures~\ref{Figure_3}(e) and (f) present the A-scan average of each B-scan depicted in Fig.~\ref{Figure_3}(a-d) with the $OPD$ scale, and with the scale calculated with $n_{\mathrm{topcoat}} \simeq 1.49$ (from 0 to 0.14~$\mu$m ) and then calculated with $n_{\mathrm{primer}} \simeq 1.84$ (blue part from from 0.14~$\mu$m). The black curves present the A-scan averages of the MIR OCT B-scan, solid line (a) and dashed line (b), while the two red curves present the A-scan averages of the NIR OCT B-scan, solid line (c) and dashed line (d). The average was made over 100 A-scans, which represent $1.146$~mm for MIR OCT and $0.29$~mm for NIR OCT. MIR OCT allows us to characterise the sample through the primer and slightly into the fine filling until around $360~\mu$m. For the NIR OCT evaluation, the signal is completely attenuated after a physical depth of $100~\mu$m due to strong scattering. In the NIR, the signal is dominated by multiple scattering, so the real OCT signal penetration depth is most likely significantly shorter \cite{Israelsen2019}. \newline

In the following figures, for a better understanding, we make the compromise to calculate the depth scale presented in the B-scans with the refractive index from the topcoat  $n_{\mathrm{topcoat}} \simeq 1.49$.

\subsection{Identification of defects by MIR OCT}

The second group of samples characterised was sample $\mathrm{n^{o}2}$, with intentional defects created at each layer (cf. Fig.~\ref{Figure_1}(c)). The process of the intentional defects is presented in section~\ref{the_blade_sample}. The punched holes subsequently undergo two processes before the layer is entirely dry: 

\begin{enumerate}
    \item [(i)] In an ideal case, the hole made after the punch doesn't move and keeps the size of the needle. Then the hole is filled with a subsequent paint layer, if it is not made in the top layer.
    \item [(ii)] In most cases, there is a more or less important reflow of wet paint after punching. This reflow could entirely obstruct the hole.
\end{enumerate}

\begin{figure}[H]
\centering\includegraphics[width=15cm]{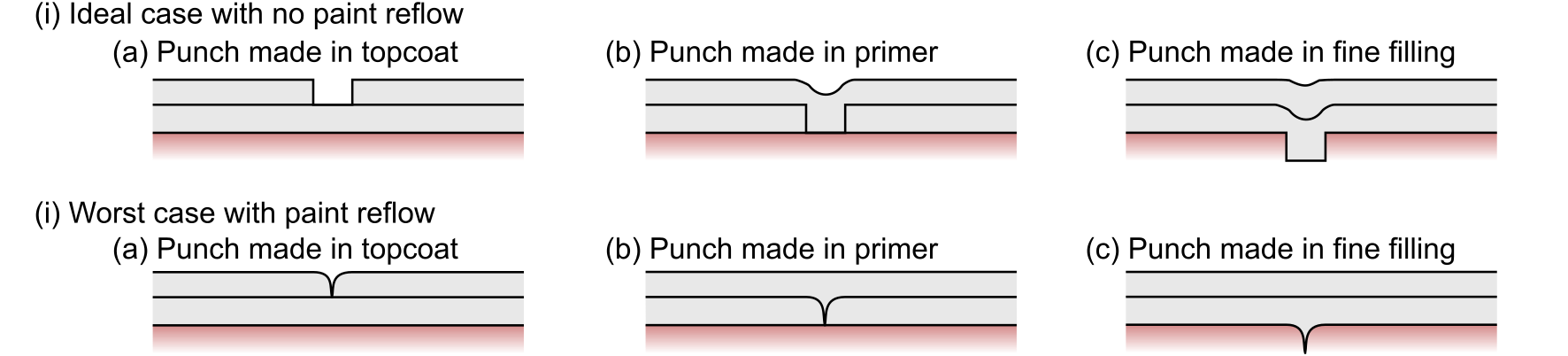}
\caption{Illustration of the impact of introducing punch defects at different layers in the production with (i) an ideal case without any reflow, and (ii) the worst case of reflow of wet paint, (a) in the topcoat, (b) in the primer, and (c) in the fine filling.}
\label{Figure_4}
\end{figure}

The resultant defects after processes (i) and (ii) are illustrated in Fig.~\ref{Figure_4}(i) and Fig.~\ref{Figure_4}(ii) for the three punched layers corresponding to (a), (b), and (c), respectively. The deeper the layer the punch is made, the less severe the disturbance in the surface of the coating is expected. In contrast, punches in the topcoat are not remedied by subsequent coating layer deposition and are most likely to persist. For this reason, punches in the topcoat will only be remedied by process (ii).

The defects identified and characterised by MIR OCT are summarised in Table~\ref{table_3}. The highest rate of identified defects was found in the topcoat until the minimum size of the needle. The relation between the identified defects and the needle size is, however, less intuitive in the subsequent layer. More defects were identified in fine filling instead of primer, these observations reflect the way how the punches were made wasn't consistent and manufacturing uncertainty. With these observations, we acknowledge that punches as small as 100~$\mu$m made in the filler are inflicting changes in the coating.\newline

\begin{table}[H]
\centering
\scalebox{0.8}{
\begin{tabular}{l|c|c|c|c|c|}
\cline{2-6}
& \multicolumn{5}{|c|}{Diameter needle} \\
\cline{2-6}
 & 3 mm       & 2 mm       & 1 mm       & 0.5 mm     & 0.1 mm     \\ \hline
\multicolumn{1}{|l|}{Topcoat}      & \checkmark & \checkmark & \checkmark & \checkmark & \checkmark \\ \hline
\multicolumn{1}{|l|}{Primer}        &            & \checkmark & \checkmark &            &            \\ \hline
\multicolumn{1}{|l|}{Fine filling}  & \checkmark & \checkmark &            & \checkmark & \checkmark \\ \hline
\multicolumn{1}{|l|}{Rough filling} &            &            &            &            &            \\ \hline
\end{tabular}}
\caption{Defects characterised by MIR OCT.}
\label{table_3}
\end{table}

\subsubsection{MIR OCT compared with X-ray CT}

\begin{figure}[H]
\centering\includegraphics[width=15cm]{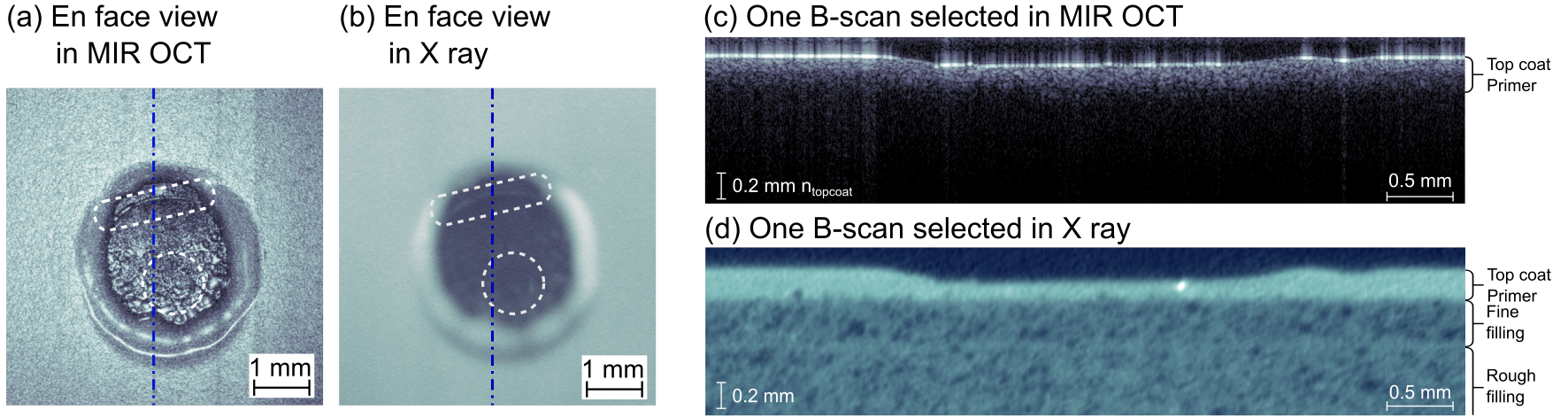}
\caption{Comparison between MIR OCT and X-ray CT measurements made of the defect made in the topcoat part with a diameter of $3$~mm. (a-b) \textit{En face} view of the surface of the sample.  (c-d) Selected B-scan in approximately the same place. The blue dashed lines in (a-b) show where the B-scans were extracted. }
\label{Figure_5}
\end{figure}

\begin{figure}[H]
\centering\includegraphics[width=15cm]{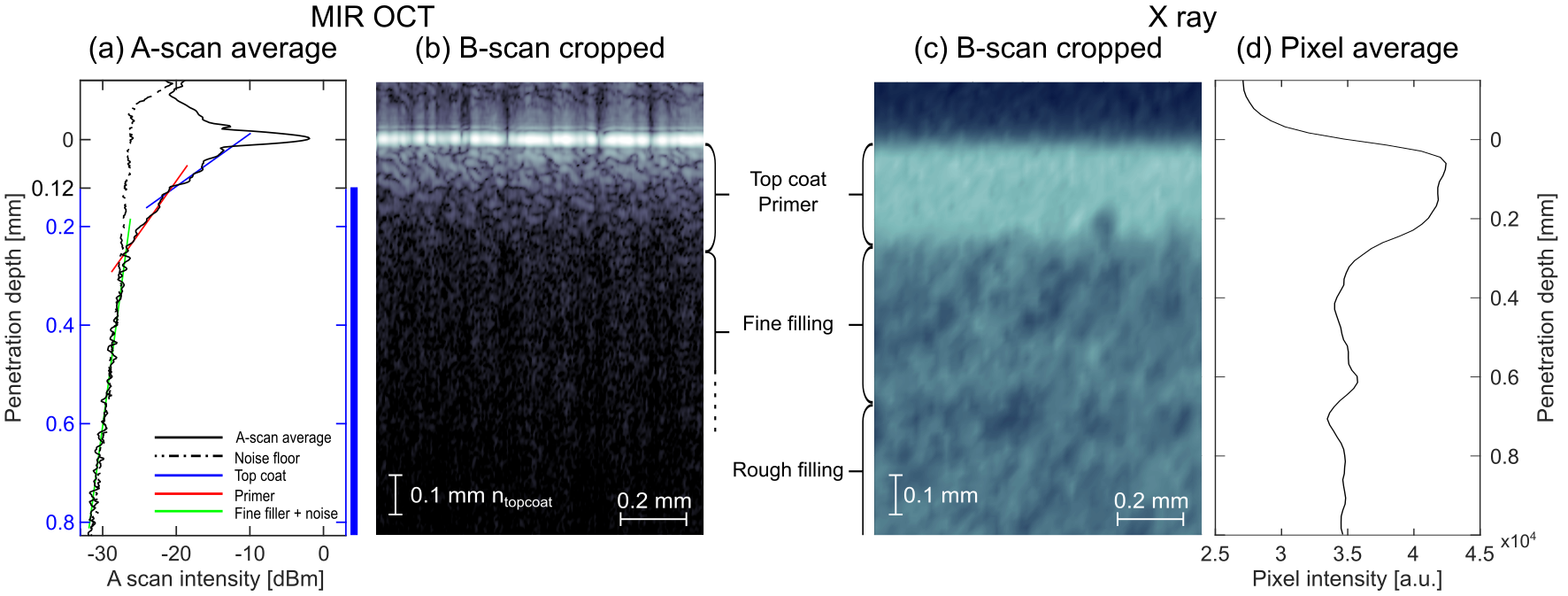}
\caption{Comparison between (b) MIR OCT B-scan and (c) X-ray CT B-scan. (a) is the A-scan average of (b) and (d) is the pixel average of (c).  For (a) the penetration depth is calculated with the $n_{\mathrm{topcoat}}$ (from 0 to 0.12~$\mu$m ) and then $n_{\mathrm{primer}}$ (blue  scale from 0.12~$\mu$m )). The solid lines blue, red, and green are the fit of topcoat, primer, and fine-filling combine with the noise floor. }
\label{Figure_6}
\end{figure}

Figures~\ref{Figure_5}-\ref{Figure_6} show comparisons between X-ray CT and OCT measurements from the sample $\mathrm{n^{o}2}$ made inside the topcoat with a $3$~mm diameter needle. The X-ray CT machine was a \textit{Zeiss Xradia 520 Versa}, with a resolution of 1-50 µm in samples of size 1-50 mm. The \textit{En face} view from OCT and X-ray CT is presented in Fig.~\ref{Figure_5}(a,b), respectively, from which it was possible  to identify the same areas on the sample, highlighted by the white dashed boxes. The resolution of X-ray CT is lower than OCT due to the large scanning area, chosen to avoid too long scanning time. Indeed, the total time for the characterisation of an area that contains 3 defects, $\sim$10$\times$1$\times$1.5 cm,  was around $6$~hours. For the OCT, the scanning time was 40 min for each defects area $\sim$5.73$\times$5.73~mm$^2$. With the resolution of MIR OCT, it is possible to detect the surface bump. \newline 

To further compare the two technologies, we selected one B-scan from MIR OCT and one B-scan from X-ray CT (cf. Fig.~\ref{Figure_5}(c,d)) approximately at the same location, shown by the dashed blue line in Fig.~\ref{Figure_5}(a,b). On both, on the right, the brackets show the different layer positions. The penetration depth is better in X-ray CT, however, the issue here is the resolution. This resolution is correlated to the sample size and so scanning area, and the larger the sample, the lower the resolution of the X-ray CT with a fixed time \cite{Mishnaevsky2021}. The resolution of OCT in depth depends only on the laser source, and the transverse resolution depends on the scanning head, and for our case, the imaging lens used. The smallest granular visuals in Fig.~\ref{Figure_5}(a) and (c), come from the presence of speckles within the measurements. \newline

To achieve a better comparison between MIR OCT and X-ray CT, we selected the first millimetre of Fig.~\ref{Figure_5}(c,d) and show this in Fig.~\ref{Figure_6}(b,c). An average A-scan is associated with the B-scan from MIR OCT (cf. Fig.~\ref{Figure_6}(a)), and a pixel average is associated with the B-scan from X-ray CT (cf. Fig.~\ref{Figure_6}(d)). For Fig.~\ref{Figure_6}(a), the spatial average was made over 0.99129~mm (173 A-scan), and for Fig.~\ref{Figure_6}(d), over 0.9984~mm (66 A-scan). In both cases, it is possible to distinguish the topcoat and the primer part, represented by the first two slopes in (a) and the first two plateaus in (d). The third slope in the MIR OCT A-scan plot is the beginning of the fine filling layer, continuing until the noise floor ((black dashed curve) in Fig.~\ref{Figure_6}) at around 0.36~mm. Beyond this point, the losses are too pronounced to detect the bottom part of the fine filling. Blue, red, and green solid lines in Fig.~\ref{Figure_6}(a) are, respectively, the fit of the topcoat, primer, and combination of fine filler plus noise.  The fitting value are:

\begin{enumerate}
\item[]  blue fit $= (-75.09\pm-9.81)\times OPD/{n_{\mathrm{topcoat}}} +(6.105 \pm 2.9)$
\item[] red fit $ =  (-47.53\pm -6.53) \times OPD/{n_{\mathrm{primer}}}+(-7.252\pm-0.89)$
\item[]  green fit $= (-4.306\pm-0.105) \times OPD+(-23.12 \pm-0.15)$
\end{enumerate}

\subsubsection{Interface extraction by MIR OCT scattering coefficient}

\begin{figure}[H]
\centering\includegraphics[width=15cm]{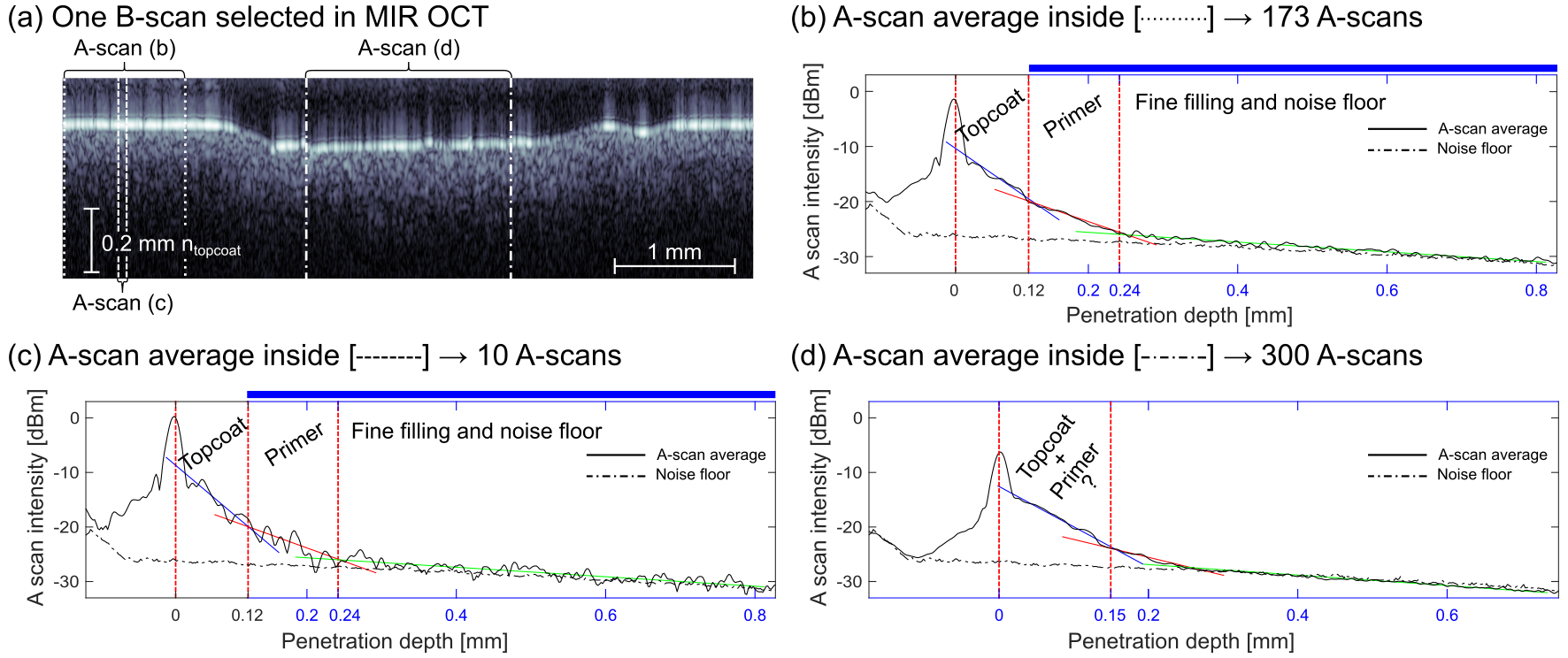}
\caption{(a) The B-scan selected is the same as in Fig.~\ref{Figure_5}, the dashed boxes show where the A-scan was extracted to calculate the average in (b-d). For (b-c) the penetration depth is calculated with the $n_{\mathrm{topcoat}}$ (from 0 to 0.12~$\mu$m ) and then $n_{\mathrm{primer}}$ (blue  scale from 0.12~$\mu$m )), and for (c) the penetration depth is calculated with $n_{\mathrm{primer}}$.}
\label{Figure_7}
\end{figure}

We conducted a study of coating thickness, topcoat, and primer on sample $\mathrm{n^{o}2}$ (cf. Fig.~\ref{Figure_7}) by tracking the bottom interface of the topcoat and the primer layer. Figure~\ref{Figure_7}(a) shows on the B-scan where are the different average A-scan regions, while Fig~\ref{Figure_7}(b-d) present the corresponding A-scan averages (black solid curves). The average parameters are for (b) 173 neighbouring A-scan  ($\sim$0.9913~mm), (c) 10 neighbouring A-scan ($\sim$0.0573~mm), and (d) 300 neighbouring A-scan ($\sim$1.719~mm). The beginning of the fine filling part can be detected up to a depth of approximately 0.36~mm, where the noise floor intersects with the A-scan averaging. \newline

The average A-scans depicted in Fig.~\ref{Figure_7}(b) and (c) present the limitation of averaging. As shown in Fig.~\ref{Figure_7}(c), the fluctuations caused by speckles are significant enough to affect the curves fitting if too few A-scans are use in the averaging. For Fig.~\ref{Figure_7}(d), the averaging was performed inside the intentional defect part. In this region we expect a certain amount of topcoat was removed after the intentional defect was made with the needle. In an ideal world, all the topcoat should be removed in this region, but the fitting slop does not correspond to the primer fit in Fig.~\ref{Figure_7}(b). It is possible that there was a non-uniform topcoat residue above the primer on the 300 A-scans. As we expect to have only primer in an ideal case, the scale for penetration depth was calculated only with $n_{\mathrm{primer}}$. 

\begin{figure}[H]
\centering\includegraphics[width=15cm]{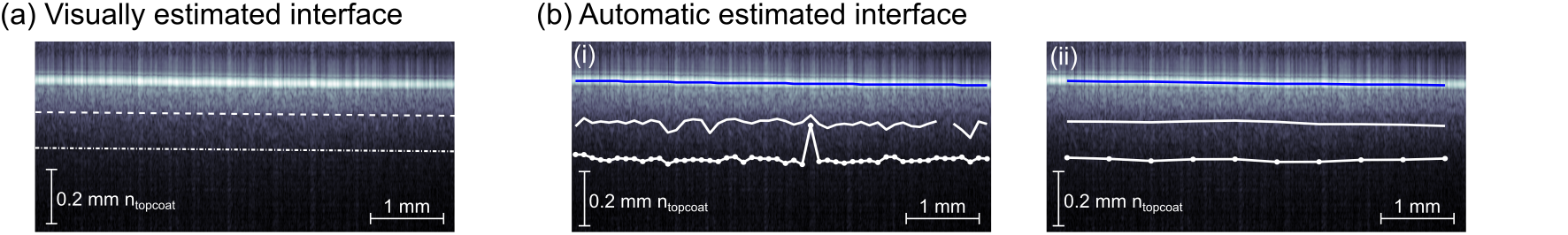}
\caption{(a) Present the estimated thickness of the topcoat and primer with visual inspection. Both lines (from top to bottom) indicate the lower boundary of the topcoat and primer. (b i) and (b ii) present the results of our algorithm for X = 20 and X = 100. The blue line represents the automatic detection of the surface, from top to bottom, while the white lines indicate the lower boundary of the topcoat and primer}
\label{Figure_8}
\end{figure}

\begin{figure}[H]
\centering\includegraphics[width=15cm]{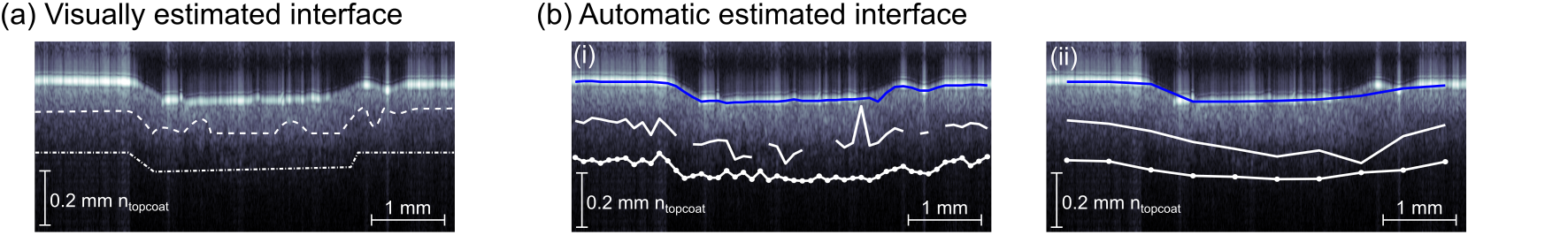}
\caption{(a) Present the estimated thickness of the topcoat and primer with visual inspection. Both lines (from top to bottom) indicate the lower boundary of the topcoat and primer. (b i) and (b ii) present the results of our algorithm for X = 20 and X = 100. The blue line represents the automatic detection of the surface, from top to bottom, while the white lines indicate the lower boundary of the topcoat and primer}
\label{Figure_9}
\end{figure}

The purpose of Fig.~\ref{Figure_8} and Fig.~\ref{Figure_9} is to demonstrate how the attenuation coefficient can be used to track the topcoat and primer thickness along a B-scan and thus across a WTB. The explanation of the algorithm is given in section~\ref{scanning_parameters}. Although estimating topcoat and primer thickness in the absence of defects is a straightforward task, the process becomes more challenging in the presence of defects. Figure~\ref{Figure_8}(a) shows the visual estimation of the bottom part of topcoat and primer outside the intentional defect, and then in Fig~\ref{Figure_8}(b), the result of our automatic interface program 20 A-scans averaged for (i) and 100 A-scans averaged for (ii).

Then we applied the program in an intentional defect region. Figure~\ref{Figure_9}(a) shows the visual estimation of the bottom part of the topcoat and primer. As seen in Fig.~\ref{Figure_7}(d), the part where the defect is can be quite challenging to estimate correctly visually. Figures~\ref{Figure_9}(b), the result of our automatic interface program 20 A-scans averaged for (i) and 100 A-scans averaged for (ii). Although the algorithm performs well at the edges of the B-scan, it encounters difficulties in distinguishing each interface in defect regions. In these areas, the slopes of the topcoat and primer layers are sometimes very similar. This issue highlights the potential benefits of integrating algorithms based on deep learning, which could significantly improve the tracking of these two layers' thicknesses.

\subsubsection{Sub-surface defect}

\begin{figure}[htbp]
\centering\includegraphics[width=15cm]{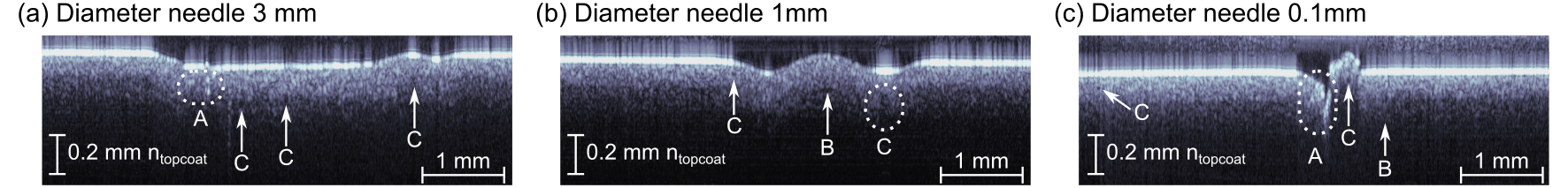}
\caption{Summary of additional defects detected. To enhance visibility, (a-c) represent an average of 5 B-scans. Defect categories found are for (a):  A and C, (b): B, and C and (c): A, B and C}
\label{Figure_10}
\end{figure}

During the manufacturing process of a WTB, some defects can occur as a result of fluctuations in the production processes. These defects may affect the general properties of the coating protecting the blade. From OCT image observations, we categorise defects as:

\begin{enumerate}
    \item [A] Inclusion - \textit{local point of high signal} (cf. Fig.~\ref{Figure_10}(a,c)).
    \item [B] Crack - \textit{elongated local region of low signal} (cf. Fig.~\ref{Figure_10}(b,c)).
    \item [C] Hole - \textit{local point of low signal} (cf. Fig.~\ref{Figure_10}(a,b,c)).
\end{enumerate}
As the OCT is sensitive to modification of refractive index, the technique is excellent for detecting this category of defects.\newline

Figure~\ref{Figure_10}, presents three typical defects that were detected in sample $\mathrm{n^{o}2}$. In Fig.~\ref{Figure_10}(a), there is the presence of bright particles inside the dashed circle (defect A) and the presence of air bubbles trapped inside the coating (defect C - see white arrows).  Figure~\ref{Figure_10}(b) shows the presence of an air bubble (defect C - left arrow), a crack (defect B - right arrow) just under the area where the intentional defect was made marked by the white arrows, and a hole (defect C - dashed circle). Due to the viscosity of the paint, the bottom part of the layer of the primer or the topcoat may have been moved. Fig.~\ref{Figure_10}(c) shows the presence of inclusion (defect A - dashed circle), air bubbles trapped inside the coating (defect C - left and middle white arrows), and the suspicion of a crack (defect B - right white arrow).\newline

\section{Discussion and conclusion}

This research highlights the capability of a MIR OCT scanner to detect internal and surface faults in WTB coatings with a penetration depth of $\sim 360~\mu$m. This depth includes topcoat, primer and beginning of the fine filling layer. Two types of samples were characterised during this study. The first sample was designed to benchmark the penetration depth of the MIR OCT system. An aluminium foil was introduced between each layer during manufacturing. Since metal easily reflects light and gives a signal above the noise floor, this was an effective method to assess whether the MIR OCT system could efficiently penetrate the coating layers, including the topcoat and primer. For better depth penetration in the future and to reach the whole coating thickness. The second sample was used to test the capability of MIR OCT to detect manufacturing defects of different categories, from µm scale to mm scale. An additional challenge with this sample involved tracking the coating layer thickness in B-scans containing fluctuations by analysing the attenuation coefficient. This opens the idea for future research of using AI-based learning techniques to support the characterisation of B-scans, especially to automatically measure the thickness of coating layers.\newline

MIR OCT images were compared to both conventional NIR OCT images and X-ray images. While MIR OCT has improved penetration performance compared to NIR OCT, it is still quite limited when compared to X-ray CT. In comparison, X-ray CT could penetrate the entire coating and GFRP of a blade, in contrary to MIR OCT where the maximum penetration depth is around 360~$\mu$m, which correspond for our case the topcoat and primer layer of the coating. However,  MIR OCT outperforms X-ray imaging in a production setting for characterising the WTB coating quality. In using an X-ray CT scanner, it is not possible to obtain a 3D microscopic map of a WTB region of interest without cutting out a piece of it. This is possible with MIR OCT, as the sample characterised doesn't need to be rotate to extract the data of the volume. Additionally, the resolution provided by the OCT measurement is better than the one provide by X-ray CT.  \newline

MIR OCT scanner has the potential of replacing the local and destructive dolly test in the production, and complementing existing methods by adding structural information on the microscopic level, which is not possible with other techniques. In addition to side-view information, it can generate a full topography map, topcoat and primer thickness maps, bulk uniformity of the coating, and, importantly, a count of the number and types of defects. It is important to minimise the number of defects inside the WTB coating, indeed, initial defects play a role in the erosion and speed up the fatigue. For example, air bubbles trapped inside the coating create stress concentration and will intensify the damage area from water droplets \cite{Faester2021, Mishnaevsky2021}.\newline

\begin{figure}[H]
\centering\includegraphics[width=15cm]{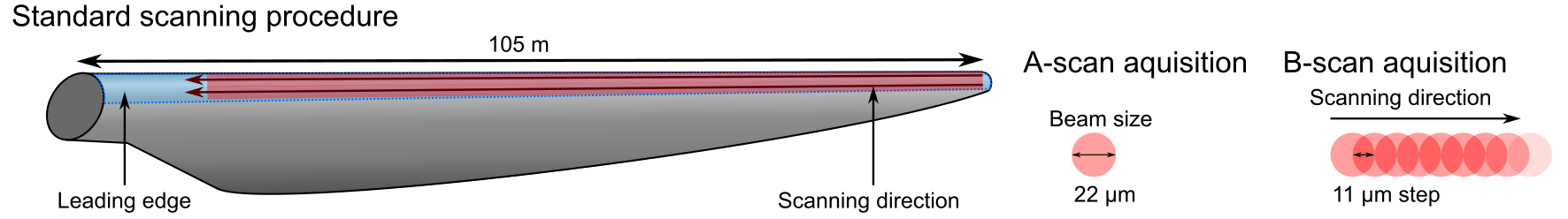}
\caption{Schematic of the proposed standard procedure for blade inspection.}
\label{Figure_11}
\end{figure}

An extraordinary insight would be obtaining the above-mentioned quality measures for a full-length WTB scan, such a scan is illustrated in Fig.~\ref{Figure_11}. Currently the scanning time is approximately 40 min for a 5.73$\times$5.73 mm$^2$ area. It is interesting to consider potential impact of a custom scanner developed WTB screening. If a production-tailored MIR OCT scanner can achieve a ten-fold improvement in speed, a scan of the full blade length of 105 m can be done within 40 minutes.

In conclusion, this study demonstrated that the MIR OCT system can track the impact of intentional defects introduced during manufacturing. The results are encouraging for using MIR OCT as a complementary tool to other NDI techniques in quality testing the full WTB volume.

\section*{Funding}

This project has received funding from Villum Fonden (2021 Villum Investigator project No.~00037822: Table-Top Synchrotrons).\newline

This project has received funding from Horizon Europe, the European Union’s Framework Programme for Research and Innovation, under Grant Agreement No.~101057404 (ZDZW), and Grant Agreement No.~101058054 (TURBO). This includes funds from UK Research and Innovation (UKRI) under the UK government’s Horizon Europe funding guarantee [grant numbers 10037822, 10042318 and 10044756] as part of the topic ID HORIZON-CL4-2021-TWIN-TRANSITION-01-02.\newline

Views and opinions expressed are however those of the authors only and do not necessarily reflect those of the European Union or UKRI. The European Union or UKRI cannot be held responsible for them

\section*{Data and Code availability}
The data presented in this study are available on request from the
corresponding author. The data are not publicly available due to significant storage requirements.

\section*{Acknowledgements}

We would like to acknowledge María Rocío Del Amor, Fernando García Torres, Natalia Lourdes Perez Garcia de la Puent, Adrian Colomer Granero and Prof. Valery Naranjo of the CVB lab, la Universitat Politècnica de València for rewarding discussions on localization and annotation of material defects observed in OCT images. We also thank the TURBO consortium for useful discussions on scope and applications of MIR OCT as an NDI technology. 

\section*{Competing interests}
The authors declare that they have no known competing financial interests or personal relationships that could have appeared to influence the work reported in this paper.

\bibliographystyle{unsrt} 
\bibliography{lapre}





\end{document}